\documentclass{article}
\usepackage{amsmath}
\usepackage{amssymb}
\usepackage{graphicx}

\input{tcilatex}

\begin{document}

\title{Logarithmic Integrals, Polylogarithmic Integrals and Euler sums}
\author{Bernard J. Laurenzi \\
%EndAName
Department of Chemistry\\
UAlbany, The State University of New York\\
1400 Washington Ave., Albany N.Y. 12222}
\date{October 24, 2010 }
\maketitle

\begin{abstract}
Relations among integrals of logarithms, polylogarithms and Euler sums are
presented. \ A unifying element being the introduction of Nielsen's
generalized polylogarithms.

Keywords: Logarithmic integrals,\ polylogarithmic integrals, Euler sums,
Nielsen's generalized polylogarithms, multiple polylogarithms
\end{abstract}

\section{Introduction}

\bigskip

Recently a number of integrals and Euler sums of interest in Physics and
Quantum Chemistry have been shown to be interrelated \cite{Bib}. When
calculating radiative corrections in Quantum Field Theory, multi-dimensional
Feynman integrals which lead to these sums are encountered. For typical
applications to Quantum Field Theory, see Cheng and Wu \cite{Cheng}. A
common theme in these works has been the presence of Neilsen's generalized
polylogarithm functions \cite{Nielsen}. \ In this article we continue these
investigations and provide a number of new identities involving integrals of
logarithms, polylogarithms, Euler sums and Neilsen functions.

\section{Polylogarithm Integrals}

\bigskip

Integrals containing the product of two polylogarithmic functions $Li_{r}(t)$
\cite{L} of orders $p$ and $q$ and having the forms ($p,q\geq1$) 
\begin{equation}
I_{+}(p,q)=\int_{0}^{1}\frac{Li_{p}(t)\,Li_{q}(t)\,dt}{t}=I_{+}(q,p),
\label{e1}
\end{equation}%
\begin{equation}
I_{-}(p,q)=\int_{0}^{1}\frac{Li_{p}(-t)\,Li_{q}(-t)\,dt}{t}=I_{-}(q,p),
\label{e2}
\end{equation}%
\begin{equation}
I_{\pm}(p,q)=\int_{0}^{1}\frac{Li_{p}(t)\,Li_{q}(-t)\,dt}{t},  \label{e3}
\end{equation}
arise in a number of physical applications \cite{Blu} and are of interest in
themselves. The integral $I_{+}(p,q)$ has been studied by Frietas \cite{F}.
We note the $p,q$ symmetry displayed in Eqs. (1) and (2). Integration by
parts for any of the integrals above allows one to obtain the following
partial difference equation 
\begin{equation}
I(p,q)+I(p+1,q-1)=R(p+1,q).  \label{e4}
\end{equation}
A more symmetric form of this difference equation being 
\begin{equation}
I(p,q-1)+I(p-1,q)=R(p,q),  \label{e5}
\end{equation}
where $I(p,q)$ stands for any of the integrals in Eqs. (1-3), and $R(p,q)$
stands for one of the three corresponding quantities%
\begin{equation*}
R_{\,+}(p,q)=Li_{p}(1)\,Li_{q}(1)=\zeta(p)\,\zeta(q),
\end{equation*}%
\begin{equation*}
R_{\,-}(p,q)=Li_{p}(-1)\,Li_{q}(-1)=(2^{1-p}-1)\zeta(p)\,(2^{1-q}-1)\zeta(q),
\end{equation*}%
\begin{equation*}
R_{\,\pm}(p,q)=Li_{p}(1)Li_{q}(-1)=\zeta(p)\,(2^{1-q}-1)\,\zeta(q),
\end{equation*}
and $\zeta(z)$ is the Riemann zeta function. We note that in the special
case where $z$ approaches $1$ that%
\begin{equation*}
\underset{z\rightarrow1}{\lim}(2^{1-z}-1)\,\zeta(z)=-\ln(2).
\end{equation*}
A set of solutions for the difference equation Eq. (5) is given by 
\begin{equation}
I(p+n,q-n)=(-1)^{n}\,\left[ \;I(p,q)-\sum_{k=0}^{n-1}(-1)^{k}R(p+k+1,q-k)%
\right] .\;  \label{e6}
\end{equation}
\textit{Proof }\textbf{:} If $p$ is replaced by $p+k$ and $q$ is replaced by 
$q-k$ in Eq. (4) we get 
\begin{equation*}
I(p+k,q-k)+I(p+k+1,q-k-1)=R(p+k+1,q-k).
\end{equation*}
Multiplying this by $(-1)^{k}$ and summing $k$ over $n$ terms produces%
\begin{equation*}
\sum_{k=0}^{n-1}(-1)^{k}\,I(p+k,q-k)+\sum_{k=0}^{n-1}(-1)^{k}%
\,I(p+k+1,q-k-1)=\sum_{k=0}^{n-1}(-1)^{k}\,R(p+k+1,q-k).
\end{equation*}
Writing out the \textit{last} term in the second sum and the\textit{\ first}
term in the first sum on the lhs of this equation produces 
\begin{equation*}
(-1)^{n-1}I(p+n,q-n)+I(p,q)=\sum_{k=0}^{n-1}(-1)^{k}\,R(p+k+1,q-k),
\end{equation*}
the remaining sums having cancelled. Rearrangement of this equation gives
the desired result.

\section{Special cases of the $I(p,q)$ integrals}

\bigskip

\subsubsection{\protect\bigskip The $I_{+}(p,q)$ and $I_{-}(p,q)$ integrals}

\paragraph{ \ \ \ \ \ The odd case $q-p=2n-1$}

In the case of $I_{+}(p,q),$ and $I_{-}(p,q)$ we note that when the
integrals' parameters $q$ and $p$ are set equal and substituted in Eq. (5)
that $I(p,p-1)=\tfrac{1}{2}R(p,p),$ or more usefully 
\begin{equation}
I(p+n,p+n-1)=\tfrac{1}{2}R(p+n,p+n),  \label{e7}
\end{equation}
i.e. the $I$ integrals can be written in closed form. This follows directly
from the $p,q$ symmetry of these integrals and Eq. (5). Consequently, the 
\emph{family} of all such integrals where $q$ and $p$ differ by an \textit{%
odd} integer can likewise be expressed in closed form. \ To see this we set $%
q=p+2n$ $-1$ in Eq. (6) and solve for $I(p,p+2n-1)$ having used Eq. (7) and
noting that the last term in the sum also contains the quantity $R(p+n,p+n).$
We find 
\begin{equation}
I(p,p+2n-1)=\tfrac{1}{2}(-1)^{n+1}R(p+n,p+n)+\sum_{k=0}^{n-2}(-1)^{k}%
\;R(p+k+1,p+2n-1-k),  \label{e8}
\end{equation}
that is, these integrals are given by a finite sum of zeta functions. Using
the $p,q$ values associated with the series defined by $p+q=5$ we have for
either $I_{+}(p,q)$ or $I_{-}(p,q),$ the examples%
\begin{equation*}
I(1,4)=-\tfrac{1}{2}\,R(3,3)+R(2,4),
\end{equation*}%
\begin{equation*}
I(2,3)=\tfrac{1}{2}\,R(3,3).
\end{equation*}

\paragraph{ \ \ \ \ \ \ The even case $q-p=2n$}

In the case where $q$ and $p$ are set equal in Eq. (4), the integrals $%
I_{+}(p,q)$ and $I_{-}(p,q)$ have the property%
\begin{equation*}
I(p+1,p-1)=R(p+1,p)-I(p,p).
\end{equation*}
All of the integrals in these families can be related to the ones which
contain the square of a single polylogarithmic function. Setting $q=p+2n$ in
Eq. (6) we get%
\begin{equation}
I(p,p+2n)=(-1)^{n}\;I(p+n,p+n)+\sum_{k=0}^{n-1}(-1)^{k}R(p+k+1,p+2n-k).
\label{e9}
\end{equation}
By way of example, for the series associated with $p+q=6$ we have for $%
I_{+}(p,q)$ or $I_{-}(p,q)$ 
\begin{equation*}
I(1,5)=I(3,3)+R(2,5)-R(3,4),
\end{equation*}%
\begin{equation*}
I(2,4)=-I(3,3)+R(3,4).
\end{equation*}

\subsubsection{\ \ \ The\ $I_{\pm}(p,q)$ integral}

\paragraph{\ \ \ \ \ \ \ \ \ The odd case}

\ \ No simple relationship such as Eq. (7) exists for $I_{\pm}(p,p-1)$,
instead there is the slightly more complicated form obtained from
Eq.\thinspace(5) where $p$ and $q$ have been set equal and $p$ has been
replaced by $p+n.$ We get 
\begin{equation*}
I_{\pm}(p+n,p+n-1)+I_{\pm}(p+n-1,p+n)=R_{\pm}(p+n,p+n).
\end{equation*}
Setting \ $q=p+2n-1$ and substituting $I_{\pm}(p+n,p+n-1)$ into Eq. (6) we
get upon writing out the last term in the sum%
\begin{equation*}
I_{\pm}(p,p+2n-1)=(-1)^{n+1}I_{\pm}(p+n-1,p+n)+\sum_{k=0}^{n-2}(-1)^{k}\;R_{%
\pm}(p+k+1,p+2n-1-k).
\end{equation*}
Here we see that knowledge of the integrals $I_{\pm}(p+n-1,p+n)$ is required
to compute $I_{\pm}(p,p+2n-1).$

\paragraph{ \ \ \ \ \ \ The even case}

Relations similar to Eq. (9) exist for the $I_{\pm}(p,q)$ integrals. If $q$
and $p$ are set equal in Eq. (4)

\begin{equation*}
I_{\pm}(p+1,p-1)=R_{\pm}(p+1,p)-I_{\pm}(p,p).
\end{equation*}

The integrals $I_{\pm}(p+1,p-1)$ are seen to be related to $I_{\pm}(p,p)$
which contains two polylogarithmic functions of order $p$ but with arguments
differing in sign. \ Setting $q=p+2n$ in Eq. (6) we get a result which has
exactly the same form as that for $I_{+}(p,q),$ or $I_{-}(p,q)$ as shown in
Eq. (9) i.e.

\begin{equation*}
I_{\pm}(p,p+2n)=(-1)^{n}\;I_{\pm}(p+n,p+n)+\sum_{k=0}^{n-1}(-1)^{k}R_{\pm
}(p+k+1,p+2n-k).
\end{equation*}

\section{Polylogarithm $I$ integrals with low order}

\bigskip

In cases with low order, some special integrals arise i.e. for $q=0,1$ we
have 
\begin{equation*}
\int_{0}^{1}\frac{Li_{p}(t)}{1+t}dt=-I_{\pm}(p,0),
\end{equation*}%
\begin{equation*}
\int_{0}^{1}\frac{Li_{p}(-t)}{1+t}dt=-I_{-}(p,0),
\end{equation*}%
\begin{equation*}
\int_{0}^{1}\frac{Li_{p}(t)-Li_{p}(1)}{1-t}dt=-I_{+}(1,p-1),
\end{equation*}%
\begin{equation*}
\int_{0}^{1}\frac{Li_{p}(-t)-Li_{p}(-1)}{1-t}dt=-I_{\pm}(1,p-1).
\end{equation*}
Integrals of the sort shown above arise in the calculation of hadronic heavy
quark production \cite{Korner}. These are examples of \textit{multiple
polylogarithms }$Li_{m_{1}},...,\,_{m_{n}}\left( x_{1},...,x_{n}\right) $ as
defined by \ \ \ \ \ \ \ \ \ \ \ \ \ \ \ \ \ \ \ \ \ \ \ \ \ \ \ \ \ \ \ \ \
\ \ \ \ \ \ \ \ \ \ \ \ \ \ \ \ \ \ \ \ \ \ \ \ \ \ \ \ \ \ \ \ \ \ \ \ \ \
\ \ \ \ \ \ \ \ \ \ \ \ \ \ \ \ \ \ \ \ \ \ \ \ \ \ \ \ \ \ \ \ \ \ \ \ \ \
\ \ \ \ \ \ \ \ \ \ \ \ \ \ \ \ \ \ \ \ \ \ \ \ \ \ \ \ \ \ \ \ \ \ \ \ \ \
\ \ \ \ \ \ \ \ \ \ \ \ \ \ \ \ \ \ \ \ \ \ \ \ \ \ \ \ \ \ \ \ \ \ \ \ \ \
\ \ \ \ \ \ \ \ \ \ \ \ \ \ \ \ \ \ \ \ \ \ \ \ \ \ \ \ \ \ \ \ \ \ \ \ \ \
\ \ \ \ \ \ \ \ \ \ \ \ \ \ \ \ \ \ \ \ \ \ \ \ \ \ \ \ \ \ \ \ \ \ \ \ \ \
\ \ \ \ \ \ \ \ \ \ \ \ \ \ \ \ \ \ \ \ \ \ \ \ \ \ \ \ \ \ \ \ \ \ \ \ \ \
\ \ \ \ \ \ \ \ \ \ \ \ \ \ \ \ \ \ \ \ \ \ \ \ \ \ \ \ \ \ \ \ \ \ \ \ \ \
\ \ \ \ \ \ \ \ \ \ \ \ \ \ \ \ \ \ \ \ \ \ \ \ \ \ \ \ \ \ \ \ \ \ \ \ \ \
\ \ \ \ \ \ \ \ \ \ \ \ \ \ \ \ \ \ \ \ \ \ \ \ \ \ \ \ \ \ \ \ \ \ \ \ \ \
\ \ \ \ \ \ \ \ \ \ \ \ \ \ \ \ \ \ \ \ \ \ \ \ \ \ \ \ \ \ \ \ \ \ \ \ \ \
\ \ \ \ \ \ \ \ \ \ \ \ \ \ \ \ \ \ \ \ \ \ \ \ \ \ \ \ \ \ \ \ \ \ \ \ \ \
\ \ \ \ \ \ \ \ \ \ \ \ \ \ \ \ \ \ \ \ \ \ \ \ \ \ \ \ \ \ \ \ \ \ \ \ \ \
\ \ \ \ \ \ \ \ \ \ \ \ \ \ \ \ \ \ \ \ \ \ \ \ \ \ \ \ \ \ \ \ \ \ \ \ \ \
\ \ \ \ \ \ \ \ \ \ \ \ \ \ \ \ \ \ \ \ \ \ \ \ \ \ \ \ \ \ \ \ \ \ \ \ \ \
\ \ \ \ \ \ \ \ \ \ \ \ \ \ \ \ \ \ \ \ \ \ \ \ \ \ \ \ \ \ \ \ \ \ \ \ \ \
\ \ \ \ \ \ \ \ \ \ \ \ \ \ \ \ \ \ \ \ \ \ \ \ \ \ \ \ \ \ \ \ \ \ \ \ \ \
\ \ \ \ \ \ \ \ \ \ \ \ \ \ \ \ \ \ \ \ \ \ \ \ \ \ \ \ \ \ \ \ \ \ \ \ \ \
\ \ \ \ \ \ \ \ \ \ \ \ \ \ \ \ \ \ \ \ \ \ \ \ \ \ \ \ \ \ \ \ \ \ \ \ \ \
\ \ \ \ \ \ \ \ \ \ \ \ \ \ \ \ \ \ \ \ \ \ \ \ \ \ \ \ \ \ \ \ \ \ \ \ \ \
\ \ \ \ \ \ \ \ \ \ \ \ \ \ \ \ \ \ \ \ \ \ \ \ \ \ \ \ \ \ \ \ \ \ \ \ \ \
\ \ \ \ \ \ \ \ \ \ \ \ \ \ \ \ \ \ \ \ \ \ \ \ \ \ \ \ \ \ \ \ \ \ \ \ \ \
\ \ \ \ \ \ \ \ \ \ \ \ \ \ \ \ \ \ \ \ \ \ \ \ \ \ \ \ \ \ \ \ \ \ \ \ \ \
\ \ \ \ \ \ \ \ \ \ \ \ \ \ \ \ \ \ \ \ \ \ \ \ \ \ \ \ \ \ \ \ \ \ \ \ \ \
\ \ \ \ \ \ \ \ \ \ \ \ \ \ \ \ \ \ \ \ \ \ \ \ \ \ \ \ \ \ \ \ \ \ \ \ \ \
\ \ \ \ \ \ \ \ \ \ \ \ \ \ \ \ \ \ \ \ \ \ \ \ \ \ \ \ \ \ \ \ \ \ \ \ \ \
\ \ \ \ \ \ \ \ \ \ \ \ \ \ \ \ \ \ \ \ \ \ \ \ \ \ \ \ \ \ \ \ \ \ \ \ \ \
\ \ \ \ \ \ \ \ \ \ \ \ \ \ \ \ \ \ \ \ \ \ \ \ \ \ \ \ \ \ \ \ \ \ \ \ \ \
\ \ \ \ \ \ \ \ \ \ \ \ \ \ \ \ \ \ \ \ \ \ \ \ \ \ \ \ \ \ \ \ \ \ \ \ \ \
\ \ \ \ \ \ \ \ \ \ \ \ \ \ \ \ \ \ \ \ \ \ \ \ \ \ \ \ \ \ \ \ \ \ \ \ \ \
\ \ \ \ \ \ \ \ \ \ \ \ \ \ \ \ \ \ \ \ \ \ \ \ \ \ \ \ \ \ \ \ \ \ \ \ \ \
\ \ \ \ \ \ \ \ \ \ \ \ \ \ \ \ \ \ \ \ \ \ \ \ \ \ \ \ \ \ \ \ \ \ \ \ \ \
\ \ \ \ \ \ \ \ \ \ \ \ \ \ \ \ \ \ \ \ \ \ \ \ \ \ \ \ \ \ \ \ \ \ \ \ \ \
\ \ \ \ \ \ \ \ \ \ \ \ \ \ \ \ \ \ \ \ \ \ \ \ \ \ \ \ \ \ \ \ \ \ \ \ \ \
\ \ \ \ \ \ \ \ \ \ \ \ \ \ \ \ \ \ \ \ \ \ \ \ \ \ \ \ \ \ \ \ \ \ \ \ \ \
\ \ \ \ \ \ \ \ \ \ \ \ \ \ \ \ \ \ \ \ \ \ \ \ \ \ \ \ \ \ \ \ \ \ \ \ \ \
\ \ \ \ \ \ \ \ \ \ \ \ \ \ \ \ \ \ \ \ \ \ \ \ \ \ \ \ \ \ \ \ \ \ \ \ \ \
\ \ \ \ \ \ \ \ \ \ \ \ \ \ \ \ \ \ \ \ \ \ \ \ \ \ \ \ \ \ \ \ \ \ \ \ \ \
\ \ \ \ \ \ \ \ \ \ \ \ \ \ \ \ \ \ \ \ \ \ \ \ \ \ \ \ \ \ \ \ \ \ \ \ \ \
\ \ \ \ \ \ \ \ \ \ \ \ \ \ \ \ \ \ \ \ \ \ \ \ \ \ \ \ \ \ \ \ \ \ \ \ \ \
\ \ \ \ \ \ \ \ \ \ \ \ \ \ \ \ \ \ \ \ \ \ \ \ \ \ \ \ \ \ \ \ \ \ \ \ \ \
\ \ \ \ \ \ \ \ \ \ \ \ \ \ \ \ \ \ \ \ \ \ \ \ \ \ \ \ \ \ \ \ \ \ \ \ \ \
\ \ \ \ \ \ \ \ \ \ \ \ \ \ \ \ \ \ \ \ \ \ \ \ \ \ \ \ \ \ \ \ \ \ \ \ \ \
\ \ \ \ \ \ \ \ \ \ \ \ \ \ \ \ \ \ \ \ \ \ \ \ \ \ \ \ \ \ \ \ \ \ \ \ \ \
\ \ \ \ \ \ \ \ \ \ \ \ \ \ \ \ \ \ \ \ \ \ \ \ \ \ \ \ \ \ \ \ \ \ \ \ \ \
\ \ \ \ \ \ \ \ \ \ \ \ \ \ \ \ \ \ \ \ \ \ \ \ \ \ \ \ \ \ \ \ \ \ \ \ \ \
\ \ \ \ \ \ \ \ \ \ \ \ \ \ \ \ \ \ \ \ \ \ \ \ \ \ \ \ \ \ \ \ \ \ \ \ \ \
\ \ \ \ \ \ \ \ \ \ \ \ \ \ \ \ \ \ \ \ \ \ \ \ \ \ \ \ \ \ \ \ \ \ \ \ \ \
\ \ \ \ \ \ \ \ \ \ \ \ \ \ \ \ \ \ \ \ \ \ \ \ \ \ \ \ \ \ \ \ \ \ \ \ \ \
\ \ \ \ \ \ \ \ \ \ \ \ \ \ \ \ \ \ \ \ \ \ \ \ \ \ \ \ \ \ \ \ \ \ \ \ \ \
\ \ \ \ \ \ \ \ \ \ \ \ \ \ \ \ \ \ \ \ \ \ \ \ \ \ \ \ \ \ \ \ \ \ \ \ \ \
\ \ \ \ \ \ \ \ \ \ \ \ \ \ \ \ \ \ \ \ \ \ \ \ \ \ \ \ \ \ \ \ \ \ \ \ \ \
\ \ \ \ \ \ \ \ \ \ \ \ \ \ \ \ \ \ \ \ \ \ \ \ \ \ \ \ \ \ \ \ \ \ \ \ \ \
\ \ \ \ \ \ \ \ \ \ \ \ \ \ \ \ \ \ \ \ \ \ \ \ \ \ \ \ \ \ \ \ \ \ \ \ \ \
\ \ \ \ \ \ \ \ \ \ \ \ \ \ \ \ \ \ \ \ \ \ \ \ \ \ \ \ \ \ \ \ \ \ \ \ \ \
\ \ \ \ \ \ \ \ \ \ \ \ \ \ \ \ \ \ \ \ \ \ \ \ \ \ \ \ \ \ \ \ \ \ \ \ \ \
\ \ \ \ \ \ \ \ \ \ \ \ \ \ \ \ \ \ \ \ \ \ \ \ \ \ \ \ \ \ \ \ \ \ \ \ \ \
\ \ \ \ \ \ \ \ \ \ \ \ \ \ \ \ \ \ \ \ \ \ \ \ \ \ \ \ \ \ \ \ \ \ \ \ \ \
\ \ \ \ \ \ \ \ \ \ \ \ \ \ \ \ \ \ \ \ \ \ \ \ \ \ \ \ \ \ \ \ \ \ \ \ \ \
\ \ \ \ \ \ \ \ \ \ \ \ \ \ \ \ \ \ \ \ \ \ \ \ \ \ \ \ \ \ \ \ \ \ \ \ \ \
\ \ \ \ \ \ \ \ \ \ \ \ \ \ \ \ \ \ \ \ \ \ \ \ \ \ \ \ \ \ \ \ \ \ \ \ \ \
\ \ \ \ \ \ \ \ \ \ \ \ \ \ \ \ \ \ \ \ \ \ \ \ \ \ \ \ \ \ \ \ \ \ \ \ \ \
\ \ \ \ \ \ \ \ \ \ \ \ \ \ \ \ \ \ \ \ \ \ \ \ \ \ \ \ \ \ \ \ \ \ \ \ \ \
\ \ \ \ \ \ \ \ \ \ \ \ \ \ \ \ \ \ \ \ \ \ \ \ \ \ \ \ \ \ \ \ \ \ \ \ \ \
\ \ \ \ \ \ \ \ \ \ \ \ \ \ \ \ \ \ \ \ \ \ \ \ \ \ \ \ \ \ \ \ \ \ \ \ \ \
\ \ \ \ \ \ \ \ \ \ \ \ \ \ \ \ \ \ \ \ \ \ \ \ \ \ \ \ \ \ \ \ \ \ \ \ \ \
\ \ \ \ \ \ \ \ \ \ \ \ \ \ \ \ \ \ \ \ \ \ \ \ \ \ \ \ \ \ \ \ \ \ \ \ \ \
\ \ \ \ \ \ \ \ \ \ \ \ \ \ \ \ \ \ \ \ \ \ \ \ \ \ \ \ \ \ \ \ \ \ \ \ \ \
\ \ \ \ \ \ \ \ \ \ \ \ \ \ \ \ \ \ \ \ \ \ \ \ \ \ \ \ \ \ \ \ \ \ \ \ \ \
\ \ \ \ \ \ \ \ \ \ \ \ \ \ \ \ \ \ \ \ \ \ \ \ \ \ \ \ \ \ \ \ \ \ \ \ \ \
\ \ \ \ \ \ \ \ \ \ \ \ \ \ \ \ \ \ \ \ \ \ \ \ \ \ \ \ \ \ \ \ \ \ \ \ \ \
\ \ \ \ \ \ \ \ \ \ \ \ \ \ \ \ \ \ \ \ \ \ \ \ \ \ \ \ \ \ \ \ \ \ \ \ \ \
\ \ \ \ \ \ \ \ \ \ \ \ \ \ \ \ \ \ \ \ \ \ \ \ \ \ \ \ \ \ \ \ \ \ \ \ \ \
\ \ \ \ \ \ \ \ \ \ \ \ \ \ \ \ \ \ \ \ \ \ \ \ \ \ \ \ \ \ \ \ \ \ \ \ \ \
\ \ \ \ \ \ \ \ \ \ \ \ \ \ \ \ \ \ \ \ \ \ \ \ \ \ \ \ \ \ \ \ \ \ \ \ \ \
\ \ \ \ \ \ \ \ \ \ \ \ \ \ \ \ \ \ \ \ \ \ \ \ \ \ \ \ \ \ \ \ \ \ \ \ \ \
\ \ \ \ \ \ \ \ \ \ \ \ \ \ \ \ \ \ \ \ \ \ \ \ \ \ \ \ \ \ \ \ \ \ \ \ \ \
\ \ \ \ \ \ \ \ \ \ \ \ \ \ \ \ \ \ \ \ \ \ \ \ \ \ \ \ \ \ \ \ \ \ \ \ \ \
\ \ \ \ \ \ \ \ \ \ \ \ \ \ \ \ \ \ \ \ \ \ \ \ \ \ \ \ \ \ \ \ \ \ \ \ \ \
\ \ \ \ \ \ \ \ \ \ \ \ \ \ \ \ \ \ \ \ \ \ \ \ \ \ \ \ \ \ \ \ \ \ \ \ \ \
\ \ \ \ \ \ \ \ \ \ \ \ \ \ \ \ \ \ \ \ \ \ \ \ \ \ \ \ \ \ \ \ \ \ \ \ \ \
\ \ \ \ \ \ \ \ \ \ \ \ \ \ \ \ \ \ \ \ \ \ \ \ \ \ \ \ \ \ \ \ \ \ \ \ \ \
\ \ \ \ \ \ \ \ \ \ \ \ \ \ \ \ \ \ \ \ \ \ \ \ \ \ \ \ \ \ \ \ \ \ \ \ \ \
\ \ \ \ \ \ \ \ \ \ \ \ \ \ \ \ \ \ \ \ \ \ \ \ \ \ \ \ \ \ \ \ \ \ \ \ \ \
\ \ \ \ \ \ \ \ \ \ \ \ \ \ \ \ \ \ \ \ \ \ \ \ \ \ \ \ \ \ \ \ \ \ \ \ \ \
\ \ \ \ \ \ \ \ \ \ \ \ \ \ \ \ \ \ \ \ \ \ \ \ \ \ \ \ \ \ \ \ 
\begin{equation*}
Li_{m_{1}},...,_{m_{n}}\left( x_{1},...,x_{n}\right)
=\sum_{k_{n}>...k_{2}>k_{1}>0}\frac{%
x_{n}^{k_{n}}...x_{2}^{k_{2}}x_{1}^{k_{1}}}{%
k_{n}^{m_{n}}...k_{2}^{m_{2}}k_{1}^{m_{1}}}.
\end{equation*}
Expressing the integrals above in these terms we have%
\begin{equation*}
\int_{0}^{1}\frac{Li_{p}(\pm t)}{1+t}dt=-Li_{p,1}(\mp\,1,-1),
\end{equation*}%
\begin{equation*}
\int_{0}^{1}\frac{Li_{p}(t)-Li_{p}(1)}{1-t}dt=Li_{p,1}(1,1)-\zeta(p+1),
\end{equation*}%
\begin{equation*}
\int_{0}^{1}\frac{Li_{p}(-t)-Li_{p}(-1)}{1-t}dt=Li_{p,1}(1,-1)+(1-2^{-p})%
\zeta(p+1).
\end{equation*}

\section{Infinite series representations for the $I$ integrals}

As noted above, for an exact calculation of the $I_{+}(p,q),$ $I_{-}(p,q)$
and $I_{\pm}(p,q)$ integrals, closed form expressions for the integrals $%
I(r,r)$ are needed for either even or odd values of $r$. \ In the case of
the quantities $I_{\pm}(p,q),$ integrals of the form $I_{\pm}(r,r+1)$ are
also required. \ As will be seen below, it is possible to express these
integrals as infinite series whose sums can be found in closed form in at
least some cases.

\bigskip

We will begin with $I_{-}(p,q)$. Expanding one of the polylogarithm
functions say $Li_{q}(-t)$ in a power series within its integrand , we get%
\begin{equation*}
I_{-}(p,q)=\sum_{k=1}^{\infty }\frac{(-1)^{k}}{k^{q}}\int_{0}^{1}Li_{p}(-t)%
\;t^{k-1}\;dt.
\end{equation*}%
The moment integrals appearing in this sum can be obtained by repeated
integration by parts and we have the general result(with empty sum for $p=1$)%
\begin{equation*}
\int_{0}^{1}Li_{p}(-t)\;t^{k-1}\;dt=\frac{(-1)^{p}}{k^{p}}[\psi (k+1)-\psi
(k/2+1)]+\sum_{\mu =2}^{p}\frac{(-1)^{p-\mu }}{k^{p+1-\mu }}(2^{1-\mu
}-1)\;\zeta (\mu ),
\end{equation*}%
where $\psi (z)$ is the Psi (Digamma) function. Boyadzhiev has computed the
moments of $Li_{p}(t)$\cite{Boyad}$.$ \ The integral $I_{-}(p,q)$ then
becomes%
\begin{equation*}
I_{-}(p,q)=(-1)^{p}\sum_{\mu =2}^{p}(-1)^{\mu }(2^{1-\mu }-1)\,\zeta (\mu
)\,(2^{\mu -p-q}-1)\,\zeta (p+q+1-\mu )
\end{equation*}%
\begin{equation*}
+\;(-1)^{p}\sum_{k=1}^{\infty }\frac{(-1)^{k}}{k^{p+q}}[\psi (k+1)-\psi
(k/2+1)].
\end{equation*}%
The infinite sums in this expression can be represented in a more convenient
form by writing out the $\psi (k/2+1)$ portion of the latter sum in terms of
its even and odd values of $k.$ We get%
\begin{equation*}
I_{-}(p,q)=(-1)^{p}\,2\,\left[ \ln (2)\,(2^{-p-q}-1)\,\zeta
(p+q)+(1-2^{-p-q-1})\,\zeta (p+q+1)\right]
\end{equation*}%
\begin{equation*}
+\,(-1)^{p}\sum_{\mu =2}^{p}(-1)^{\mu }(2^{1-\mu }-1)\,\zeta (\mu )\,(2^{\mu
-p-q}-1)\,\zeta (p+q+1-\mu )
\end{equation*}%
\begin{equation*}
+\,\,(-1)^{p}\,\left[ \sum_{k=1}^{\infty }\frac{(-1)^{k}[\psi (k+1)+\gamma ]%
}{k^{p+q}}-\sum_{k=1}^{\infty }\frac{[\psi (k+1)+\gamma ]}{(2k)^{p+q}}%
+\sum_{k=0}^{\infty }\frac{[\psi (k+1/2)-\psi (1/2)]}{(2k+1)^{p+q}}\right] ,
\end{equation*}%
where $\gamma $ is Euler's constant. In a similar way the integrals $%
I_{+}(p,q)$ and $I_{\pm }(p,q)$ are given by%
\begin{equation*}
I_{+}(p,q)=(-1)^{p}\sum_{\mu =2}^{p}(-1)^{\mu }\,\zeta (\mu )\,\zeta
(p+q+1-\mu )+(-1)^{p+1}\sum_{k=1}^{\infty }\frac{1}{k^{p+q}}[\psi
(k+1)+\gamma ],
\end{equation*}%
\begin{equation*}
I_{\pm }(p,q)=(-1)^{p}\sum_{\mu =2}^{p}(-1)^{\mu }\,\zeta (\mu )\,(2^{\mu
-p-q}-1)\,\zeta (p+q+1-\mu )+(-1)^{p+1}\sum_{k=1}^{\infty }\frac{(-1)^{k}}{%
k^{p+q}}[\psi (k+1)+\gamma ].
\end{equation*}%
Although not manifestly obvious, the $p,q$ symmetry is preserved in the
infinite sum relations for $I_{+}(p,q)$ and $I_{-}(p,q)$.

\section{Closed forms for the infinite series}

\bigskip

In the expressions for the $I$ integrals given above, four infinite series
containing Psi functions occur. We will show below that it is possible to
give closed form expressions for two of them whereas in the remaining two,
closed forms can be given only for \textit{even} values of $p+q$.

\bigskip

We begin with a discussion of the Euler sum $S_{+}(r)$ of order $r$ i.e.%
\begin{equation*}
S_{+}(r)=\sum_{k=1}^{\infty}\frac{1}{k^{r}}[\psi(k+1)+\gamma].
\end{equation*}
A closed form expression for this sum where $r\geq2$ had originally been
given by Euler i.e. 
\begin{equation*}
S_{+}(r)=\tfrac{1}{2}(r+2)\,\zeta(r+1)-\tfrac{1}{2}\sum_{\mu=1}^{r-2}\zeta
(\mu+1)\,\zeta(r-\mu)\,.
\end{equation*}
In the case of the \textit{alternating }Euler sum $S_{-}(r)$ i.e. 
\begin{equation}
S_{-}(r)=\sum_{k=1}^{\infty}\frac{(-1)^{k}}{k^{r}}[\psi(k+1)+\gamma],
\label{e10}
\end{equation}
the situation is more complicated. Rewriting the series in Eq. (10) as sums
corresponding to the even and odd values of $k$ respectively, we find 
\begin{equation*}
S_{-}(r)=\sum_{k=1}^{\infty}\frac{[\psi(2k+1)+\gamma]}{(2k)^{r}}-\sum
_{k=0}^{\infty}\frac{[\psi(2k+2)+\gamma]}{(2k+1)^{r}}.
\end{equation*}
Each of the Psi functions in these sums can be replaced by functions of
lower argument using the Psi function's duplication formulas. We get%
\begin{equation*}
S_{-}(r)=\tfrac{1}{2}\sum_{k=1}^{\infty}\frac{\psi(k+1/2)}{(2k)^{r}}+\tfrac {%
1}{2}\sum_{k=1}^{\infty}\frac{\psi(k+1)}{(2k)^{r}}+(\gamma+\ln(2))\,2^{-r}\,%
\zeta(r)
\end{equation*}

\begin{equation*}
-\tfrac{1}{2}\sum_{k=0}^{\infty }\frac{\psi (k+1)}{(2k+1)^{r}}-\tfrac{1}{2}%
\sum_{k=0}^{\infty }\frac{\psi (k+1/2)}{(2k+1)^{r}}-(\gamma +\ln
(2))\,(1-2^{-r})\,\zeta (r)-(1-2^{-r-1})\,\zeta (r+1).
\end{equation*}%
This expression can be simplified with the help of four known sums which we
call the Jordan sums $\mathcal{J}_{1}(r)$, $\mathcal{J}_{2}(r)$ ; the
Milgram sum $\mathcal{M}(r),$ and the sum $\mathcal{C}(r)$ where 
\begin{subequations}
\label{11}
\begin{align}
\mathcal{J}_{1}(r)& =\tfrac{1}{2}\sum_{k=0}^{\infty }\frac{\left[ \psi
(k+1/2)-\psi (1/2)\right] }{(2k+1)^{r}},  \label{11a} \\
\mathcal{J}_{2}(r)& =\tfrac{1}{2}\sum_{k=1}^{\infty }\frac{\left[ \psi
(k+1/2)-\psi (1/2)\right] }{(2k)^{r}},  \label{11b} \\
\mathcal{M}(r)& =\tfrac{1}{2}\sum_{k=0}^{\infty }\frac{\left[ \psi
(k+1)+\gamma \right] }{(2k+1)^{r}},  \label{11c} \\
\mathcal{C}(r)& =\tfrac{1}{2}\sum_{k=1}^{\infty }\frac{\left[ \psi
(k+1)+\gamma \right] }{(2k)^{r}}=\tfrac{1}{2^{r+1}}\,S_{+}(r).  \label{11d}
\end{align}%
The sum $S_{-}(r)$ can then be rewritten as 
\end{subequations}
\begin{equation*}
S_{-}(r)=\mathcal{J}_{2}(r)-\mathcal{J}_{1}(r)+\mathcal{C}(r)-\mathcal{M}%
(r)-(1-2^{-r-1})\,\zeta (r+1).
\end{equation*}%
Closed form expressions for the sums $\mathcal{J}_{1}(r)$ and $\mathcal{J}%
_{2}(r)$ have been given by Jordan \cite{J} only for even values of $r$ i.e.

\begin{equation*}
\mathcal{J}_{1}(2n)=-\tfrac{1}{2}(1-2^{-2n-1})\,\zeta(2n+1)+%
\ln(2)(1-2^{-2n})\,\zeta(2n)
\end{equation*}

\begin{equation*}
-\;2^{-2n-1}\sum_{\mu =1}^{n-1}(2^{2\mu }-1)\,\zeta (2\mu )\,\zeta
(2n+1-2\mu ),\qquad (n\geq 1),
\end{equation*}%
and 
\begin{equation*}
\mathcal{J}_{2}(2n)=\tfrac{1}{2}(1-2^{-2n-1})\,\zeta (2n+1)-\sum_{\mu
=1}^{n-1}\,(2^{-2\mu }-2^{-2n-1})\,\zeta (2\mu )\,\zeta (2n+1-2\mu ).
\end{equation*}%
The sum $\mathcal{M(}r\mathcal{)}$ has been given by Milgram \cite{M} for $%
r\geq 2$ in closed form (with empty sum for $r\leq 2)$ as%
\begin{align*}
\mathcal{M}(r)& =\tfrac{1}{2}r\,(1-2^{-r-1})\,\zeta (r+1)-\ln
(2)\,(1-2^{-r})\,\zeta (r) \\
& -\,\tfrac{1}{2(r-1)}\sum_{\mu =0}^{r-3}(\mu +1)(2^{\mu +2}-1)\,\zeta (\mu
+2)(2^{-\mu -1}-2^{-r})\,\zeta (r-1-\mu ).
\end{align*}%
This Milgram sum can be simplified considerably because of the reoccurrence
of the zeta products within the sum. \ We get 
\begin{align*}
\mathcal{M}(r)& =\tfrac{1}{2}r\,(1-2^{-r-1})\,\zeta (r+1)-\ln
(2)\,(1-2^{-r})\,\zeta (r) \\
& -\tfrac{1}{2}\sum_{\mu =0}^{(r-4)/2}(2^{\mu +2}-1)\,\zeta (\mu +2)(2^{-\mu
-1}-2^{-r})\,\zeta (r-1-\mu ),\text{ \ for even }r
\end{align*}%
and%
\begin{align*}
\mathcal{M}(r)& =\tfrac{1}{2}r\,(1-2^{-r-1})\,\zeta (r+1)-\ln
(2)\,(1-2^{-r})\,\zeta (r)-\tfrac{1}{2}\left[ (1-2^{-\left( \tfrac{r+1}{2}%
\right) })\zeta (\tfrac{r+1}{2})\right] ^{2} \\
& -\tfrac{1}{2}\sum_{\mu =0}^{(r-5)/2}(2^{\mu +2}-1)\,\zeta (\mu +2)(2^{-\mu
-1}-2^{-r})\,\zeta (r-1-\mu ),\text{ \ for odd }r
\end{align*}%
The closed form expression for the sum $\mathcal{C}(r),$ has already been
given above as a multiple of $S_{+}(r)$.\bigskip 

Finally, the values of the $I$ integrals can be written in terms of these
sums as%
\begin{equation*}
I_{\,+}(p,q)=(-1)^{p}\sum_{\mu=2}^{p}(-1)^{\mu}\,\zeta(\mu)\,\zeta
(p+q+1-\mu)+(-1)^{p+1}S_{+}(p+q),
\end{equation*}%
\begin{equation*}
I_{\,\pm}(p,q)=(-1)^{p}\sum_{\mu=2}^{p}(-1)^{\mu}\,\zeta(\mu)\,(2^{\mu
-p-q}-1)\,\zeta(p+q+1-\mu)+(-1)^{p+1}S_{-}(p+q),
\end{equation*}%
\begin{equation*}
I_{\,-}(p,q)=(-1)^{p}\,2\,\left[ \ln(2)\,(2^{-p-q}-1)\,\zeta
(p+q)+(1-2^{-p-q-1})\,\zeta(p+q+1)\right]
\end{equation*}%
\begin{equation*}
+\,(-1)^{p}\sum_{\mu=2}^{p}(-1)^{\mu}(2^{1-\mu}-1)\,\zeta(\mu)\,(2^{\mu
-p-q}-1)\,\zeta(p+q+1-\mu)
\end{equation*}%
\begin{equation*}
+\,(-1)^{p}\left[ S_{-}(p+q)-2\,\mathcal{C}(p+q)+2\,\mathcal{J}_{1}(p+q)%
\right] .
\end{equation*}

To summarize the situation, we have found that closed form expressions for $%
I_{+}(p,q)$ and $I_{-}(p,q)$ when $q=p+2n-1$ are given by Eq. (8). In the
cases where $q=p+2n$ the sums $S_{+}(2p+2n)$ and $S_{-}(2p+2n),$ which are
known in closed form, are required for the calculation of $I_{+}(p,q)$ and $%
I_{-}(p,q$).\bigskip

In the case of $I_{\,\pm}(p,q)$ where $q=p+2n-1$ the sum $S_{-}(2p+2n-1)$ is
required but this in not generally known in closed form. \ When $q=p+2n$
where $S_{-}(2p+2n)$ is required, the integral $I_{\,\pm}(p,2p+2n)$ is
calculable in closed form.

\subsection{Connection with Nielsen's polylogarithm functions}

The alternating Euler sums $S_{-}(r)$ and the Jordan sums $\mathcal{J}%
_{1}(r) $ and $\mathcal{J}_{2}(r)$ can be expressed in terms of Nielsen's
generalized polylogarithm functions \cite{kandl} $\mathcal{S}_{n,\,p}\,(z)$
i.e.%
\begin{equation*}
\mathcal{S}_{n,\,p}\,(z)=\frac{(-1)^{n+p-1}}{(n-1)!\,p!}\int_{0}^{1}\frac{%
\ln^{n-1}(x)\,\ln^{p}(1-z\,x)}{x}dx,
\end{equation*}
and the three special cases $\mathcal{S}_{n,1}\,(z)=Li_{n+1}(z)$, $\mathcal{S%
}_{n,\,p}\,(1)=s_{n,\,\,p}$ and $\mathcal{S}_{n,\,p}\,(-1)=\widetilde{\sigma}%
\,_{n,\,p}$ \ where the latter two integrals are given by

\begin{equation*}
s_{n,\,p}=\frac{(-1)^{n+p-1}}{(n-1)!\,p!}\int_{0}^{1}\frac{%
\ln^{n-1}(x)\,\ln^{p}(1-x)}{x}dx=s_{p,\,n},
\end{equation*}
and%
\begin{equation*}
\widetilde{\sigma}\,_{n,\,p}=\frac{(-1)^{n+p-1}}{(n-1)!\,p!}\int_{0}^{1}%
\frac{\ln^{n-1}(x)\,\ln^{p}(1+x)}{x}dx.
\end{equation*}
In the case of $S_{-}(r)$ Coffey \cite{C} has given the following integral
representation for $S_{-}(r)$%
\begin{equation*}
S_{-}(r)=\frac{(-1)^{r}}{\Gamma(r)}\int_{0}^{1}\left( \frac{1}{x}-\frac{1}{%
1+x}\right) \ln^{r-1}(x)\ln(1+x)\,dx,
\end{equation*}
which can be rewritten in terms of the polylogarithm function and Nielsen's
generalized polylogarithm function $\mathcal{S}_{n,\,p}\,(z)$ i.e. 
\begin{equation*}
S_{-}(r)=Li_{r+1}\,(-1)+\mathcal{S}_{r-1,\,2}\,(-1)=(2^{-r}-1)\zeta (r+1)+%
\widetilde{\sigma}_{r-1,\,2}.
\end{equation*}
Both Lewin and Kolbig \cite{kandl} have commented on their inability to
express $\mathcal{S}_{n,p}\,(-1)$ in closed form for even values of$\ n$
thus leaving the search for a closed form expressions for $%
S_{-}(2n+1)=(2^{-2n-1})\,\zeta(2n+2)+\widetilde{\sigma}_{2n\,,2}$ as an open
question.\bigskip

It is interesting that closed form expressions for the sums $\mathcal{J}%
_{1}(r)$, and $\mathcal{J}_{2}(r)$ are not generally known for odd values of 
$r$ except in the case of the two series 
\begin{equation*}
\tfrac{1}{2}\sum_{k=0}^{\infty }\frac{\left[ \psi (k+1/2)-\psi (1/2)\right] 
}{(2k+1)^{3}}=\frac{23}{5760}\pi ^{4}+\frac{\pi ^{2}}{24}\ln ^{2}(2)-\frac{1%
}{24}\ln ^{4}(2)-Li_{4}(1/2),
\end{equation*}%
\begin{equation}
\tfrac{1}{2}\sum_{k=1}^{\infty }\frac{\left[ \psi (k+1/2)-\psi (1/2)\right] 
}{(2k)^{3}}=\frac{7}{8}\ln (2)\,\zeta (3)-\frac{53}{5760}\pi ^{4}-\frac{\pi
^{2}}{24}\ln ^{2}(2)+\frac{1}{24}\ln ^{4}(2)+\,Li_{4}(1/2),  \label{e12}
\end{equation}%
which have been obtained (cf. Appendix A1) from work by Coffey \cite{c2}.
The methods used to obtain these closed form expressions do not seem to
apply for any other odd powers of $r$ for the sums in question. Using the
sums above, $S_{-}(3)$ is then given in closed form by

\begin{equation*}
S_{-}(3)=\frac{7}{4}\ln(2)\,\zeta(3)-\frac{11}{360}\pi^{4}-\frac{\pi^{2}}{12}%
\ln^{2}(2)+\frac{1}{12}\ln^{4}(2)+2\,Li_{4}(1/2).
\end{equation*}

It is also possible to find integral representation (cf. Appendix A1) for
the sums $\mathcal{J}_{1}(r)$, $\mathcal{J}_{2}(r)$ and $\mathcal{C}(r)$
which are in turn expressible in terms of the quantities $s_{n,\,m}$ and $%
\tilde{\sigma }_{n,\,m}$. \ Finally the $I$ integrals written in terms of
the Nielsen integrals are given by%
\begin{equation*}
(-1)^{p}\,I_{\,+}(p,q)=\sum_{\mu=2}^{p}(-1)^{\mu}\,\zeta(\mu)\,\zeta
(p+q+1-\mu)-\zeta(p+q+1)-s_{p+q-1,\,2},
\end{equation*}

\begin{equation*}
(-1)^{p}\,I_{\,\pm}(p,q)=\sum_{\mu=2}^{p}(-1)^{\mu}\,\zeta(\mu)\,(2^{\mu
-p-q}-1)\,\zeta(p+q+1-\mu)-(2^{-p-q}-1)\zeta(p+q+1)-\tilde{\sigma}%
_{p+q-1,\,2},
\end{equation*}

\begin{align*}
(-1)^{p}\,I_{\,-}(p,q) &
=2\ln(2)\,(2^{-p-q}-1)\,\zeta(p+q)+2(1-2^{-p-q-1})\,\zeta(p+q+1) \\
& +\sum_{\mu=2}^{p}(-1)^{\mu}(2^{1-\mu}-1)\,\zeta(\mu)\,(2^{\mu
-p-q}-1)\,\zeta(p+q+1-\mu) \\
& +(1-2^{-p-q})\,s_{p+q-1,\,2}-\zeta(p+q+1)-2\,\mathcal{M}(p+q).
\end{align*}
Written in these terms, the integrals $I_{+}$ and $I_{-\text{ \ }}$are seen
to be expressible in closed form in all cases whereas closed form
expressions for $I_{\pm}$ are obtainable for even $\ p+q$. \ Only in the
case where $p+q=3$ are closed form expressions for $I_{\pm}$ obtainable by
the methods used here.

\section{Logarithmic Integrals\protect\bigskip}

\paragraph{\protect\bigskip The $i(n,m)$ integrals}

The integrals 
\begin{equation*}
i(n,m)=\int_{0}^{1}\ln^{n}(x)\ln^{m}(1-x)\,dx=i(m,n),
\end{equation*}
are related to Nielsen's generalized polylogarithms $s_{n,\,\,p}$.
Integration by parts yields the relations%
\begin{equation*}
\frac{i(n,m)}{n!\,m!}=(-1)^{m+n}-\,\sum_{\mu=0}^{m-1}(-1)^{\mu}\;\frac{%
i(n-1,m-\mu)}{(n-1)!\,(m-\mu)!}-(-1)^{m+n}\sum_{\mu=0}^{m-1}s_{n,\,m-\mu}\;.
\end{equation*}
If the quantities $i^{\ast}(n,m)=(-1)^{m+n}\,i(n,m)/m!\,n!$ are introduced
into the equation above we get%
\begin{equation}
i^{\ast}(n,m)=1+\sum_{\mu=0}^{m-1}i^{\ast}(n-1,m-\mu)-\sum_{%
\mu=0}^{m-1}s_{n,\,m-\mu}\;.  \label{e13}
\end{equation}
Writing out the lhs of this equation for the case $i^{\ast}(n,m-1)$ and
subtracting it from Eq. (13) we get Pascal's triangular, three term partial
difference equation%
\begin{equation}
i^{\ast}(n,m)-i^{\ast}(n,m-1)-i^{\ast}(n-1,m)=-\,s_{n,\,m}.  \label{e14}
\end{equation}
This can be solved for $i^{\ast}(n,m)$ in closed form using the Laplace
method of generating functions \cite{jpde} with the initial condition $%
i^{\ast }(0,m)=1.$ The final result is 
\begin{equation*}
(-1)^{m+n}\,i(n,m)=(m+n)!-m\cdot n\cdot(m+n-2)!\,\zeta(2)-m\cdot n!\sum
_{\nu=2}^{n}\frac{(n-\nu+m-1)!}{(n-\nu)!}\zeta(\nu+1)
\end{equation*}%
\begin{equation*}
-n\cdot m!\;\sum_{\mu=2}^{m}\frac{(n-\mu+m-1)!}{(n-\mu)!}\zeta(\mu
+1)-m!\,n!\,\sum_{\mu=2}^{m}\sum_{\nu=2}^{n}\frac{(n-\nu+m-\mu)!}{%
(n-\nu)!\,(m-\mu)!}\,s_{\nu,\,\mu}.
\end{equation*}

The integrals $s_{n,\,p}$ have been given by Kolbig \cite{K2} in closed
form. \ Alternatively, they can be quickly generated with a computer algebra
system such as Maple or Mathematica using the Kolbig relation%
\begin{equation*}
s_{n,\,p}=\frac{(-1)^{n+p-1}}{(n-1)!\,p!}\left[ \frac{\partial^{n+p-1}}{%
\partial\beta^{n-1}\partial\alpha^{p}}\left\{ \frac{1}{\beta}\frac{%
\Gamma(1+\alpha)\,\Gamma(1+\beta)}{\Gamma(1+\alpha+\beta)}\right\} \right]
_{\alpha=\beta=0}.
\end{equation*}
Using this expression, a number of the $i(n,m)$ integrals have been obtained
and are displayed in the table below.

\begin{center}
\bigskip Table 1. $i(n,m),1\leq n,m\leq 3$
\end{center}

\begin{equation*}
\begin{tabular}{|c|c|c|c|c|c|c|c|c|}
\hline
$i(n,m)$ & $(-1)^{m+n}(m+n)!$ & $\pi ^{2}$ & $\pi ^{4}$ & $\pi ^{6}$ & $%
\zeta (3)$ & $\zeta (5)$ & $\pi ^{2}\zeta (3)$ & $\zeta ^{2}(3)$ \\ \hline
$i_{1,1}$ & $2$ & $-1/6$ &  &  &  &  &  &  \\ \hline
$i_{1,\,2}$ & $-6$ & $1/3$ &  &  & $2$ &  &  &  \\ \hline
$i_{1,\,3}$ & $24$ & $-1$ & $-1/15$ &  & $-6$ &  &  &  \\ \hline
$i_{2,\,2}$ & $24$ & $-4/3$ & $-1/90$ &  & $-12$ &  &  &  \\ \hline
$i_{2,\,3}$ & $-120$ & $6$ & $1/6$ &  &  & $24$ & $-2$ &  \\ \hline
$i_{3,\,3}$ & $720$ & $-36$ & $-1$ & $-23/420$ & $-216$ & $-144$ & $12$ & $%
36 $ \\ \hline
\end{tabular}%
\end{equation*}%
An alternate and more direct way to compute the $i(n,m)$ integrals is due to
a suggestion by Milgram \cite{Milsuggest} i.e. noting the relation%
\begin{equation*}
\beta (\nu +1,\mu +1)=\int_{0}^{1}x^{\mu }(1-x)^{\nu }dx,
\end{equation*}%
differentiation $n$ times with respect to $\mu $ and $m$ times with respect
to $\nu $ and evaluated at $\mu =\nu =0$ give%
\begin{equation*}
\left[ \frac{\partial ^{m+n}}{\partial \nu ^{m}\partial \mu ^{n}}\beta (\nu
+1,\mu +1)\right] _{\mu =\nu =0}=i(m,n)
\end{equation*}

As a final remark, we note that the three term partial difference equation
Eq. (14) can alternately be viewed as a way to compute the $s_{n,\,\,m}$
integrals given independently computed values of the $i(n,m)$ integrals.

\bigskip

\subsubsection{The $h(n,m)$ integrals}

\bigskip

In a similar way the integrals defined by%
\begin{equation*}
h(n,m)=\int_{0}^{1}\ln ^{n}(x)\ln ^{m}(1+x)\,dx,
\end{equation*}%
can be integrated by parts to give%
\begin{equation*}
\frac{h(n,m)}{n!\,m!}=(-1)^{m+n}+\sum_{\mu =0}^{m-1}(-1)^{n-1+m-\mu }\;\frac{%
h(n-1,m-\mu )}{(n-1)!\,(m-\mu )!}+(-1)^{m+n}\sum_{\mu =0}^{m-1}\widetilde{%
\sigma }_{n,\,\,m-\mu }\;,
\end{equation*}%
which upon introducing the quantities $h^{\ast
}(n,m)=(-1)^{m+n}\,h(n,m)/n!\,m!$\ gives%
\begin{equation*}
h^{\ast }(n,m)=1+\sum_{\mu =0}^{m-1}h^{\ast }(n-1,m-\mu )+\sum_{\mu =0}^{m-1}%
\widetilde{\sigma }_{n,\,\,m-\mu }\;.
\end{equation*}%
As shown above, a three term partial difference equation for the quantities $%
h^{\ast }(n,m)$ can similarly be obtained i.e.%
\begin{equation*}
h^{\ast }(n,m)-h^{\ast }(n,m-1)-h^{\ast }(n-1,m)=\widetilde{\sigma }%
_{n,\,\,m}.
\end{equation*}%
In this case, the method of generating functions together with the initial
conditions $h(0,m)=-1+2\,e_{m}(-\ln (2)),$ where $e_{m}(x)$ is the truncated
exponential function, gives the solution%
\begin{equation*}
(-1)^{n+m}h(n,m)=-(n+m)!+2n!\left[ -\ln (2)\right] ^{m}%
\,_{2}F_{2}([-m,n+1],[\;],1/\ln (2))
\end{equation*}%
\begin{equation*}
+m!\,n!\,\sum_{\mu =1}^{m}\sum_{\nu =1}^{n}\frac{(n-\nu +m-\mu )!}{(n-\nu
)!\,(m-\mu )!}\,\,\widetilde{\sigma }_{\nu ,\,\,\mu },
\end{equation*}%
and where $_{2}F_{2}$ is a Gauss generalized hypergeometric function which
reduces to a polynomial for the values of the parameters $n,m$ encountered
here. We have%
\begin{equation*}
_{2}F_{2}([-m,n+1],[\;],1/\ln (2))=\frac{m!}{n!}\sum_{k=0}^{m}\frac{(n+k)!}{%
(m-k)!}\frac{[-1/\ln (2)]^{k}}{k!},
\end{equation*}%
so that 
\begin{equation*}
(-1)^{n+m}h(n,m)=(n+m)!+2\,m!\sum_{\mu =1}^{m}\frac{(m+n-\mu )!}{(m-\mu )!}%
\frac{[-\ln (2)]^{\mu }}{\mu !}
\end{equation*}%
\begin{equation*}
+m!\,n!\,\sum_{\mu =1}^{m}\sum_{\nu =1}^{n}\frac{(n-\nu +m-\mu )!}{(n-\nu
)!\,(m-\mu )!}\,\,\widetilde{\sigma }_{\nu ,\,\,\mu }.
\end{equation*}%
Values for the first few $h(n,m)$ integrals are given below.%
\begin{equation*}
h(1,1)=2-2\ln (2)-\frac{\pi ^{2}}{12},
\end{equation*}%
\begin{equation*}
\underline{h(n,m),\;\text{\ }n+m=3}
\end{equation*}
\begin{eqnarray*}
h(1,2) &=&6-\frac{\pi ^{2}}{6}+\frac{\zeta (3)}{4}-8\ln (2)+2\ln ^{2}(2), \\
h(2,1) &=&-6+\frac{\pi ^{2}}{6}+\frac{3}{2}\zeta (3)+4\ln (2),
\end{eqnarray*}%
\begin{equation*}
\underline{h(n,m),\;\text{\ }n+m=4}
\end{equation*}%
\begin{eqnarray*}
h(1,3) &=&24-36\ln (2)+12\ln ^{2}(2)-2\ln ^{3}(2)+\ln ^{4}(2)/4-\pi
^{2}/2-\pi ^{4}/15 \\
&&+3\,\zeta (3)/4-\pi ^{2}\ln ^{2}(2)/4+21\ln (2)\,\zeta
(3)/4+6\,Li_{4}(1/2), \\
h(2,2) &=&24-24\ln (2)+4\ln ^{2}(2)+\ln ^{4}(2)/3-2\pi ^{2}/3-\pi ^{4}/12 \\
&&-5\,\zeta (3)/2-\pi ^{2}\ln ^{2}(2)/3+7\ln (2)\,\zeta (3)+8\,Li_{4}(1/2),
\\
h(3,1) &=&24-12\ln (2)-7\ln ^{4}(2)/120-\pi ^{2}/2-9\,\zeta (3)/2,
\end{eqnarray*}%
\begin{equation*}
\underline{h(n,m),\;\text{\ }n+m=5}
\end{equation*}%
\begin{eqnarray*}
h(1,4) &=&-120+192\ln (2)-72\ln ^{2}(2)+16\ln ^{3}(2)-3\ln ^{4}(2)+2\pi
^{2}+4\pi ^{4}/15 \\
&&-3\,\zeta (3)+\pi ^{2}\ln ^{2}(2)-21\ln (2)\,\zeta (3)-24\,Li_{4}(1/2), \\
h(2,3) &=&-120+144\ln (2)-36\ln ^{2}(2)+4\ln ^{3}(2)-3\ln ^{4}(2)/2+4\ln
^{5}(2)/5+3\pi ^{2} \\
&&+23\pi ^{4}/60+[6-\pi ^{2}-63\ln (2)/2+21\ln ^{2}(2)/2]\,\zeta
(3)-99\,\zeta (5)/8 \\
&&+(3/2-2\ln (2)/3)\,\pi ^{2}\ln ^{2}(2)+[24\ln
(2)\,-36]\,Li_{4}(1/2)+24\,Li_{5}(1/2), \\
h(3,2) &=&-120+96\ln (2)-12\ln ^{2}(2)-\ln ^{4}(2)+3\pi ^{2}+11\pi ^{4}/30 \\
&&+[33/2-\pi ^{2}-21\ln (2)]\,\zeta (3)+87\,\zeta (5)/8+\pi ^{2}\ln
^{2}(2)-24\,Li_{4}(1/2), \\
h(4,1) &=&-120+48\ln (2)+2\pi ^{2}+7\pi ^{4}/30+18\zeta (3)+45\,\zeta (5)/2,
\end{eqnarray*}%
\bigskip

\subsection{Relations between the $s_{n,\,p}$ and $\widetilde{\protect\sigma 
}\,_{n,\,p}$ integrals}

The integrals $s_{n,\,p}$ and $\widetilde{\sigma}_{n,\,p}$ \ are linearly
related. These relations are worth examining since they shed some light on
the problem of finding closed form expressions for the even values of $n$ in
the $\widetilde{\sigma}_{n,\,p}$ integrals. \ To show this we note that

\begin{equation*}
\left[ \tfrac{\partial^{n+m}}{\partial\mu^{n}\partial\nu^{m}}\left( \tfrac{%
\Gamma(\mu)\Gamma(\nu)}{\Gamma(\mu+\nu)}\right) \right] _{\mu
=0\,,\nu=1}=\int _{0}^{1}\tfrac{\ln^{n}(x/1+x))\ln^{m}(1/(1+x))}{x(1+x)}%
dx+\int _{0}^{1}\tfrac{\ln^{m}(x/1+x))\ln^{n}(1/(1+x))}{(1+x)}dx{\small ,}
\end{equation*}
where the integral representation for the Beta function%
\begin{equation*}
\beta(\mu,\nu)=\frac{\Gamma(\mu)\Gamma(\nu)}{\Gamma(\mu+\nu)}=\int_{0}^{1}%
\frac{x^{\mu-1}+x^{\nu-1}}{(1+x)^{\mu+\nu}}\,dx,
\end{equation*}
has been used. \ Expanding the logarithm terms in the integrals above,
together with the relations 
\begin{equation*}
\int_{0}^{1}\frac{\ln^{q}(x)\ln^{p}(1+x)}{(1+x)}dx=-\left( \tfrac{q}{p+1}%
\right) \int_{0}^{1}\frac{\ln^{q-1}(x)\ln^{p+1}(1+x)}{x}\,dx,
\end{equation*}
we find%
\begin{equation*}
(-1)^{n+m}s_{n+1,\,m}=\sum_{k=0}^{n}(-1)^{k}\tbinom{n+m-1-k}{m-1}\widetilde{%
\sigma}_{k+1,\,n+m-k}+\sum_{k=0}^{m-1}(-1)^{k}\tbinom{n+m-1-k}{n}\widetilde{%
\sigma}_{k+1,\,n+m-k},
\end{equation*}
or in its more symmetric form%
\begin{equation}
(-1)^{n+m}s_{n,\,\,m}=\sum_{k=1}^{n}(-1)^{k}\tbinom{n+m-1-k}{m-1}\widetilde{%
\sigma}_{k,\,n+m-k}+\sum_{k=1}^{m}(-1)^{k}\tbinom{n+m-1-k}{n-1}\widetilde{%
\sigma}_{k,\,n+m-k}.  \label{e15}
\end{equation}
As previously pointed out by Kolbig \cite{Kno}, we see that for a given
value of $m+n$ there are (for $m+n$ $\geq6)$ more of\ the$\ \widetilde{\sigma%
}_{\nu,\,\,\mu}$ integrals than there are equations, given the structure of
the relations given above in Eq. (15). \ Closed form values for some of the $%
\widetilde{\sigma}_{\nu,\,\,\mu}$ integrals are given in the tables below%
\begin{align*}
\widetilde{\sigma}_{1,\,1} & =-\pi^{2}/12 \\
\widetilde{\sigma}_{1,\,2} & =\zeta(3)/8 \\
\widetilde{\sigma}_{2,\,1} & =-3\zeta(3)/4
\end{align*}
\bigskip

\begin{center}
Table 2. $\widetilde{\sigma}_{\nu,\,\mu},\,\nu+\mu=4$

\begin{equation*}
\begin{tabular}{|c|c|c|c|c|c|}
\hline
$\widetilde{\sigma}_{\nu,\,\mu}$ & $\pi^{4}$ & $\pi^{2}\ln^{2}(2)$ & $\ln
^{4}(2)$ & $\ln(2)\zeta(3)$ & $Li_{4}(1/2)$ \\ \hline
$\widetilde{\sigma}_{1,\,3}$ & $-1/90$ & $-1/24$ & $1/24$ & $7/8$ & $1$ \\ 
\hline
$\widetilde{\sigma}_{2,\,2}$ & $-1/48$ & $-1/12$ & $1/12$ & $7/4$ & $2$ \\ 
\hline
$\widetilde{\sigma}_{3,1}$ & $-7/720$ &  &  &  &  \\ \hline
\end{tabular}%
\end{equation*}

\bigskip

Table 3. $\widetilde{\sigma}_{\nu,\,\mu},\,\nu+\mu=5$
\end{center}

\begin{equation*}
\begin{tabular}{|c|c|c|c|c|c|c|c|}
\hline
$\widetilde{\sigma}_{\nu,\,\mu}$ & $\pi^{2}\zeta(3)$ & $\zeta(5)$ & ln$%
^{2}(2)\zeta(3)$ & $\pi^{2}$ln$^{3}(2)$ & ln$^{5}(2)$ & $\ln(2)Li_{4}(1/2) $
& $Li_{5}(1/2)$ \\ \hline
$\widetilde{\sigma}_{1,\,4}$ &  & $1$ & $-7/16$ & $1/36$ & $-1/30$ & $-1$ & $%
-1$ \\ \hline
$\widetilde{\sigma}_{2,\,3}$ & $1/12$ & $33/32$ & $-7/8$ & $1/18$ & $-1/15$
& $-2$ & $-2$ \\ \hline
$\widetilde{\sigma}_{3,\,2}$ & $1/12$ & $-29/32$ &  &  &  &  &  \\ \hline
$\widetilde{\sigma}_{4,1}$ &  & $-15/16$ &  &  &  &  &  \\ \hline
\end{tabular}%
\end{equation*}

\bigskip

In the case where $\nu+\mu=6$, it is not possible to obtain closed form
expressions for the corresponding $\widetilde{\sigma}_{\nu,\,\mu}$ values
using these methods. \ However the following relations can be found.

\begin{align*}
\widetilde{\sigma}_{1,\,5} & =-\tfrac{1}{945}\pi^{6}-\tfrac{1}{96}%
\pi^{2}\ln^{4}(2)+\tfrac{1}{72}\ln^{6}(2)+\tfrac{7}{28}\ln^{3}(2)\,\zeta(3)
\\
& +\tfrac{1}{2}\ln^{2}(2)Li_{4}(1/2)+\ln(2)Li_{5}(1/2)+Li_{6}(1/2), \\
& \\
2\,\widetilde{\sigma}_{2,\,4}-\widetilde{\sigma}_{4,\,2} & =-\tfrac {53}{%
15120}\pi^{6}-\tfrac{1}{24}\pi^{2}\ln^{4}(2)+\tfrac{1}{18}\ln ^{6}(2)+\tfrac{%
7}{12}\ln^{3}(2)\,\zeta(3) \\
& -\tfrac{1}{2}\zeta^{2}(3)+2\ln^{2}(2)Li_{4}(1/2)+4%
\ln(2)Li_{5}(1/2)+4Li_{6}(1/2), \\
& \\
2\,\widetilde{\sigma}_{3,\,3}-3\,\widetilde{\sigma}_{4,\,2} & =\tfrac {1}{%
1512}\pi^{6}-\tfrac{1}{2}\zeta^{2}(3), \\
& \\
\widetilde{\sigma}_{5,\,1} & =-\tfrac{31}{30240}\pi^{6}.
\end{align*}

\bigskip

\appendix  

\section{\protect\bigskip Appendix}

\subsection{Jordan sums with odd order $r$}

It is possible to express the Jordan sum $\mathcal{J}_{1}(2n+1)$ with odd
argument as the integral%
\begin{equation*}
\mathcal{J}_{1}(2n+1)=\frac{1}{4(2n)!}\int_{0}^{1}\ln^{2n}(x)\ln\left( \frac{%
1+x}{1-x}\right) \left[ \frac{1}{1-x}-\frac{1}{1+x}\right] \,dx.
\end{equation*}
This can be done with the help of an integral representation \cite{NBS2} for
the Psi function in the numerator and the usual integral representation for
the denominator within the summand of Eq. (11a) i.e.

\begin{equation*}
\frac{\psi(k+1/2)-\psi(1/2)}{2(2k+1)^{2n+1}}=\frac{1}{2(2n)!}%
\int_{0}^{\infty }u^{2n}\exp(-[2k+1]\,u)\,du\,\int_{0}^{1}\frac{(1-t^{k})}{%
\surd t\;(1-t)}\,dt.
\end{equation*}
Summation over $k$ followed by integration over $t$ produces the desired
integral representation for $\mathcal{J}_{1}(2n+1).$

In the case where $n=1,$ $\mathcal{J}_{1}(3)$ contains the integrals

\begin{equation*}
\int_{0}^{1}\frac{\ln^{2}(x)\ln(1-x)}{1-x}dx=-\,\frac{\pi^{4}}{180},
\end{equation*}%
\begin{equation*}
\int_{0}^{1}\frac{\ln^{2}(x)\ln(1+x)}{1-x}dx=\tfrac{7}{2}\ln(2)\,\zeta (3)-%
\tfrac{19}{720}\pi^{4},
\end{equation*}%
\begin{equation*}
\int_{0}^{1}\frac{\ln^{2}(x)\ln(1-x)}{1+x}dx=\tfrac{\pi^{4}}{90}+\tfrac {%
\pi^{2}}{6}\ln^{2}(2)-\tfrac{1}{6}\ln^{4}(2)-4Li_{4}(1/2),
\end{equation*}%
\begin{equation}
\int_{0}^{1}\frac{\ln^{2}(x)\ln(1+x)}{1+x}dx=4Li_{4}(1/2)-\tfrac{\pi^{2}}{24}%
-\tfrac{\pi^{2}}{6}\ln^{2}(2)+\tfrac{1}{6}\ln^{4}(2)+\tfrac{7}{2}%
\ln(2)\,\zeta(3),  \label{e16}
\end{equation}
which have been given by Gastmann \cite{Gastmann}, Lewin \cite{L2} and
Coffey \cite{C}. \ These have been used to produce the sums given above in
Eq. (12).\bigskip

Using the same methods employed above in Eq. (11b), the integral
representation for $\mathcal{J}_{2}(2n+1)$ is given by

\begin{equation*}
\mathcal{J}_{2}(2n+1)=\tfrac{1}{4(2n)!}\int_{0}^{1}\ln^{2n}(x)\,\ln\left( 
\frac{1+x}{1-x}\right) \left[ \frac{1}{1-x}+\frac{1}{1+x}\right] \,dx,
\end{equation*}
where we note the difference in signs within the integrands of the $\mathcal{%
J}_{1}$and $\mathcal{J}_{2}$ integral representations. \ This integral upon
expansion can be rewritten as 
\begin{equation*}
\mathcal{J}_{2}(2n+1)=\tfrac{1}{4(2n)!}\int_{0}^{1}\ln^{2n}(x)\,\left[ \frac{%
\ln(1+x)}{(1+x)}-\frac{\ln(1-x)}{(1-x)}-\frac{d}{dx}\ln(1+x)\ln (1-x)\right]
dx.
\end{equation*}
The first two integrals are related to Nielsen's generalized polylogarithms $%
s_{n,\,p}$ and $\widetilde{\sigma}_{n,p}$ and we get 
\begin{equation*}
\mathcal{J}_{2}(2n+1)=\tfrac{1}{4}[s_{2n,2}+\sigma_{2n,2}]-\tfrac{1}{4(2n)!}%
\int_{0}^{1}\ln^{2n}(x)\,\left[ \frac{d}{dx}\ln(1+x)\ln(1-x)\right] dx.
\end{equation*}
The remaining integration within the $\mathcal{J}_{2}$ integral
representation can be carried out using the expansion%
\begin{equation*}
-\frac{d}{dx}\ln(1+x)\ln(1-x)=\sum_{k=1}^{\infty}\left[ 2\ln(2)-\psi
(k)+\psi(k+1/2)\right] \,x^{2k-1}.
\end{equation*}
We get 
\begin{equation*}
\tfrac{1}{4(2n)!}\int_{0}^{1}\ln^{2n}(x)\,\left[ -\,\frac{d}{dx}\ln
(1+x)\ln(1-x)\right] dx=\tfrac{1}{4}\sum_{k=1}^{\infty}\frac{\left[
2\ln(2)-\psi(k)+\psi(k+1/2)\right] }{(2k)^{2n+1}}.
\end{equation*}
The rhs of this equation can be written in terms of the sums defined above as%
\begin{equation*}
\tfrac{1}{2}\left[ \mathcal{J}_{2}(2n+1)-\mathcal{C}(2n+1)+2^{-2n-2}%
\zeta(2n+2)\right] .
\end{equation*}
The combined integrals give $\mathcal{J}_{2}(2n+1)$ as%
\begin{equation*}
\mathcal{J}_{2}(2n+1)=\tfrac{1}{2}[s_{2n,\,2}+\widetilde{\sigma}_{2n,\,2}]-%
\mathcal{C}(2n+1)+2^{-2n-2}\zeta(2n+2).
\end{equation*}

In a similar way, the sum $\mathcal{J}_{1}(2n+1)$ can be expressed in terms
of Nielsen's polylogarithms as follows. \ Adding the integral
representations for $\mathcal{J}_{1}$ and $\mathcal{J}_{2}$ we get%
\begin{equation*}
\mathcal{J}_{1}(2n+1)+\mathcal{J}_{2}(2n+1)=\tfrac{1}{2(2n)!}%
\int_{0}^{1}\ln^{2n}(x)\ln(\frac{1+x}{1-x})\frac{dx}{1-x}.
\end{equation*}
Expanding the second logarithmic term results in an integral which is in
part expressible as $s_{2n,\,2}$ i.e.%
\begin{equation*}
\mathcal{J}_{1}(2n+1)+\mathcal{J}_{2}(2n+1)=\tfrac{1}{2}\,s_{2n,\,2}+\tfrac {%
1}{2(2n)!}\int_{0}^{1}\frac{\ln^{2n}(x)\ln(1+x)}{1-x}dx.
\end{equation*}
Expansion of $\ln(1+x)/(1-x)$ in a power series gives%
\begin{equation*}
\frac{\ln(1+x)}{(1-x)}=\sum_{k=1}^{\infty}\left[ \ln(2)+(-1)^{k+1}\left\{
\psi(k/2+1)-\psi(k+1)+\ln(2)\right\} \right] \,x^{k}.
\end{equation*}
The remaining integral is then given by%
\begin{equation*}
\frac{1}{2(2n)!}\int_{0}^{1}\frac{\ln^{2n}(x)\ln(1+x)}{1-x}dx=\frac{1}{2}%
\sum_{k=1}^{\infty}\frac{\left[ \ln(2)+(-1)^{k+1}\left\{ \psi(k/2+1)-\psi
(k+1)+\ln(2)\right\} \right] }{(k+1)^{2n+1}}.
\end{equation*}
As in the case of $\mathcal{J}_{2}(2n+1)$ these infinite summations are
expressible in terms of sums defined above and we get%
\begin{equation*}
\mathcal{J}_{1}(2n+1)=\tfrac{1}{2}[s_{2n,\,2}-\widetilde{\sigma}_{2n,\,2}]-%
\mathcal{M}(2n+1).
\end{equation*}
More generally, using the same methods one can show for arbitrary $r$ that 
\begin{align*}
\mathcal{J}_{1}(r) & =\tfrac{1}{2}[s_{r-1,\,2}-\widetilde{\sigma}_{r-1,\,2}]-%
\mathcal{M}(r), \\
\mathcal{J}_{2}(r) & =\tfrac{1}{2}[s_{r-1,\,2}+\widetilde{\sigma}_{r-1,\,2}]-%
\mathcal{C}(r)+2^{-r-1}\zeta(r+1).
\end{align*}
\bigskip

The sum $\mathcal{C}(r)$ can also be written as the integral%
\begin{equation*}
\mathcal{C}(r)=\frac{(-1)^{r}}{2^{r+1}\,(r-1)!}\int_{0}^{1}\frac{\ln
^{r-1}(x)\log (1-x)}{x(1-x)}dx,
\end{equation*}%
using the methods used above in connection with the Jordan sums. \ This
integral can then be expressed in terms of the Nielsen integrals as 
\begin{equation*}
\mathcal{C}(r)=\tfrac{1}{2^{r+1}}\left( s_{r,1}+s_{r-1,\,2}\right) =\tfrac{1%
}{2^{r+1}}\left[ \zeta (r+1)+s_{r-1,\,2}\right] .
\end{equation*}%
As a result of the latter simplified expression, $\mathcal{J}_{2}(r)$ can be
written as%
\begin{equation*}
\mathcal{J}_{2}(r)=\tfrac{1}{2}[(1-2^{-r})\,s_{r-1,\,2}+\widetilde{\sigma }%
_{r-1,\,2}].
\end{equation*}%
In the case of the Milgram sum $\mathcal{M}(r),$ it does not appear to be
possible to express it in terms of the Nielsen integrals.

\subsection{Approximate values for the alternating Euler sums}

Since it does not appear to be possible to express the Jordan sums of odd
order as closed form expressions and thus the corresponding alternating
Euler sums, we have sought approximations for the later quantities.
Beginning with the relation 
\begin{equation}
\int_{0}^{1}\frac{\left[ Li_{p}(-t)-Li_{p}(-1)\right] }{1-t}\,dt=-\,\int
_{0}^{1}\frac{Li_{p}(-t)\,\ln(1-t)}{t}\,dt,  \label{e17}
\end{equation}
expansion of $Li_{p}(-t)\,$on the rhs of this equation followed by
integration with respect to $t$ gives the alternating series $S_{-}(p)$ i.e.%
\begin{equation}
\int_{0}^{1}\frac{\left[ Li_{p}(-t)-Li_{p}(-1)\right] }{1-t}\,dt=\sum
_{k=1}^{\infty}\frac{(-1)^{k}}{k^{p}}\left[ \psi(k+1)+\gamma\right]
=S_{-}(p).  \label{e18}
\end{equation}
If within the lhs of Eq. (17) $Li_{p}(-t)$ is expanded in powers of $t-1$,
we get the relation 
\begin{equation}
\int_{0}^{1}\frac{Li_{p}(-t)-Li_{p}(-1)}{1-t}dt=-\sum_{k=1}^{\infty}\frac{1}{%
k!}\int_{0}^{1}(t-1)^{k-1}dt\sum_{j=1}^{k}S_{k}^{(j)}\;(2^{1+j-p}-1)\;%
\zeta(p-j),  \label{e19}
\end{equation}
with the help of formulas for the derivatives \cite{Wolf} of the
polylogarithm functions i.e.

\begin{equation*}
\left[ \frac{d^{k}Li_{p}(-t)}{dt^{k}}\right] _{t=1}=%
\sum_{j=1}^{k}S_{k}^{(j)}\;(2^{1+j-p}-1)\;\zeta(p-j),
\end{equation*}
and where $S_{k}^{(j)}$ are the Stirling numbers of the first kind \cite{NBS}%
. Integrating over $t$ in Eq. (19) we get%
\begin{equation*}
S_{-}(p)=\sum_{k=1}^{\infty}\frac{(-1)^{k+1}}{k\cdot k!}%
\sum_{j=1}^{k}S_{k}^{(j)}\;(2^{1+j-p}-1)\zeta(p-j).
\end{equation*}
If the sum on the right-hand side of this equation is truncated to order $%
k=k_{t}$ , we obtain a value of the \textit{alternating }Euler sum which can
be made accurate to any desired number of decimals places. For example
taking $p=5$ and $k_{t}=10,$ an approximate value of the\textit{\ alternating%
} sum is

\begin{equation*}
S_{-}(5)\thickapprox\sum_{k=1}^{10}\frac{(-1)^{k+1}}{k\cdot k!}%
\sum_{j=1}^{k}S_{k}^{(j)}\;(2^{j-4}-1)\,\zeta(5-j),
\end{equation*}
which when written out explicitly takes on the form 
\begin{equation*}
S_{-}(5)\thickapprox-\;\tfrac{24387227}{1741824000}-\,\tfrac{358039}{11197440%
}\pi^{2}-\tfrac{1968329}{130636800}\pi^{4}+\tfrac{2152309}{3456000}%
\,\zeta(3)+\tfrac{1874237}{14515200}\ln(2),
\end{equation*}
an expression which is accurate to nine decimal places.\bigskip

It is hoped that future work in this area will produce exact or more
accurate values of the Euler alternating sums.

\end{document}